\documentclass[journal,draftclsnofoot, onecolumn]{IEEEtran}

\usepackage{amsmath, graphics,amssymb,epsfig,subfigure,epstopdf}
\usepackage[utf8]{inputenc}
\usepackage{multirow, cite}
\usepackage{xcolor,colortbl, soul}
\usepackage{caption}
\usepackage{hyperref}
\usepackage[percent]{overpic}

\usepackage{tabularx,booktabs}

\definecolor{Gray}{gray}{0.85}
\definecolor{LightCyan}{rgb}{0.88,1,1}
\definecolor{LightBlue}{rgb}{0.75,0.936,1.00}
\sethlcolor{LightBlue}
\newcolumntype{a}{>{\columncolor{Gray}}c}
\newcolumntype{b}{>{\columncolor{white}}c}
\definecolor{lightgray}{gray}{0.95}
\usepackage{bm}
\begin{document}

\title{\huge Deep Reinforcement Learning for Interference Management in UAV-based 3D Networks:\\ Potentials  and Challenges} 
\author{Mojtaba Vaezi, \textit{Senior Member, IEEE,}  Xingqin Lin \textit{Senior Member, IEEE,}  Hongliang Zhang,  \textit{Member, IEEE,}  Walid Saad,  \textit{Fellow, IEEE,}  and H. Vincent Poor,  \textit{Life Fellow, IEEE}\vspace{-3mm}
 \thanks{M. Vaezi is with the Department of Electrical and Computer Engineering, Villanova University, PA, USA (e-mail: mvaezi@villanova.edu).
 \newline \indent Xingqin Lin is with NVIDIA, USA (e-mail:
 xingqinl@nvidia.com).
 \newline \indent H.  Zhang is with the School of Electronics, Peking University, Beijing, China (e-mail: hongliang.zhang@pku.edu.cn).
 \newline \indent W. Saad is with the Bradley Department of Electrical and
 Computer Engineering, Virginia Tech, Arlington, VA 22203 USA (e-mail: walids@vt.edu).
 \newline \indent H. V. Poor is with the Department of Electrical and Computer Engineering, Princeton University, Princeton, NJ, USA (e-mail: poor@princeton.edu).
  }
}
\maketitle


\begin{abstract}

Modern cellular networks are multi-cell and use universal frequency reuse  to maximize \textit{spectral efficiency}. This results in high inter-cell interference. 
 This problem is growing as cellular networks become 
 three-dimensional with the adoption of unmanned aerial vehicles (UAVs). This is because the strength and number of interference links rapidly increase due to the line-of-sight channels in UAV communications.
 Existing interference management solutions
 need each transmitter to know the channel information of interfering signals, rendering them impractical due to excessive signaling overhead.
 In this paper, we propose leveraging deep reinforcement learning for interference management to tackle this shortcoming. In particular, we show that interference can still be effectively mitigated even without knowing its channel information. We then discuss novel  approaches to scale the algorithms with linear/sublinear complexity and decentralize them using multi-agent reinforcement learning. By harnessing interference, the proposed solutions 
  enable the continued growth of civilian UAVs. 
\end{abstract}

\section{Introduction}
\label{sec:intro}

 Unmanned aerial vehicles
(UAVs), better known as drones, have found a wide variety of applications in the past years.
UAVs are inherently mobile and rely on wireless connectivity to support their operational needs (command and control communication) and mission-related payload data transmission. 
The deployment of UAVs over cellular networks is transforming today's terrestrial mobile networks into three-dimensional (3D) networks.
3D networks are fundamentally different from classical 2D networks. 
This is because the high altitude of UAVs creates a radio environment where  UAV-associated channels    
are line-of-sight dominant  \cite{amorim2017radio}, thus causing strong air-to-ground and ground-to-air
interference.
Meanwhile, the mobility of UAVs increases the system's dynamics and brings forth new challenges such as delay and Doppler shifts.

The strong air-ground
interference 
severely
limits the cellular network capacity and adversely affects coexisting aerial and
terrestrial nodes. 
The 3rd Generation Partnership Project (3GPP) lists uplink/downlink interference detection and mitigation
as a major challenge of communications for UAVs \cite{3GPP}.
Simulations and field trials indicate that mobility-related performance (e.g.,
handover failure) of aerial users is worse than
terrestrial users \cite{3GPP} due to high interference. These challenges are further exacerbated by the fact that interference can 
make basic tasks like maintaining a connection to the cellular network more difficult for 
UAVs. Therefore, the wide-scale deployment of UAVs will only be possible if interference can be properly mitigated or harnessed \cite{antennagain,Challita2019Deep}.

Inter-cell interference management in cellular networks is a long-standing problem and has been extensively studied for classical 2D networks. 
Interference alignment \cite{jafar2011interference} 
and coordinated multipoint (CoMP) \cite{irmer2011coordinated} are two prominent  examples of such efforts. In theory, these techniques are much more efficient than \textit{time-division multiple access (TDMA)}, but they have some shortcomings that prevent them from being used in real-world networks. Notably, they require a global knowledge of \textit{channel 
	state information} (CSI) which is not practically feasible, and their implementation incurs large signaling overhead and  needs tight synchronization between cooperating nodes, which reduces their effectiveness in practice. 
Given these limitations, 
\textit{orthogonalizing} the resources (time/frequency/beam) and  \textit{treating interference as noise} remains the most common solution for addressing co-channel interference in practice.

\begin{figure}[ht!] 
	\centering
	\includegraphics[width=.45\columnwidth]{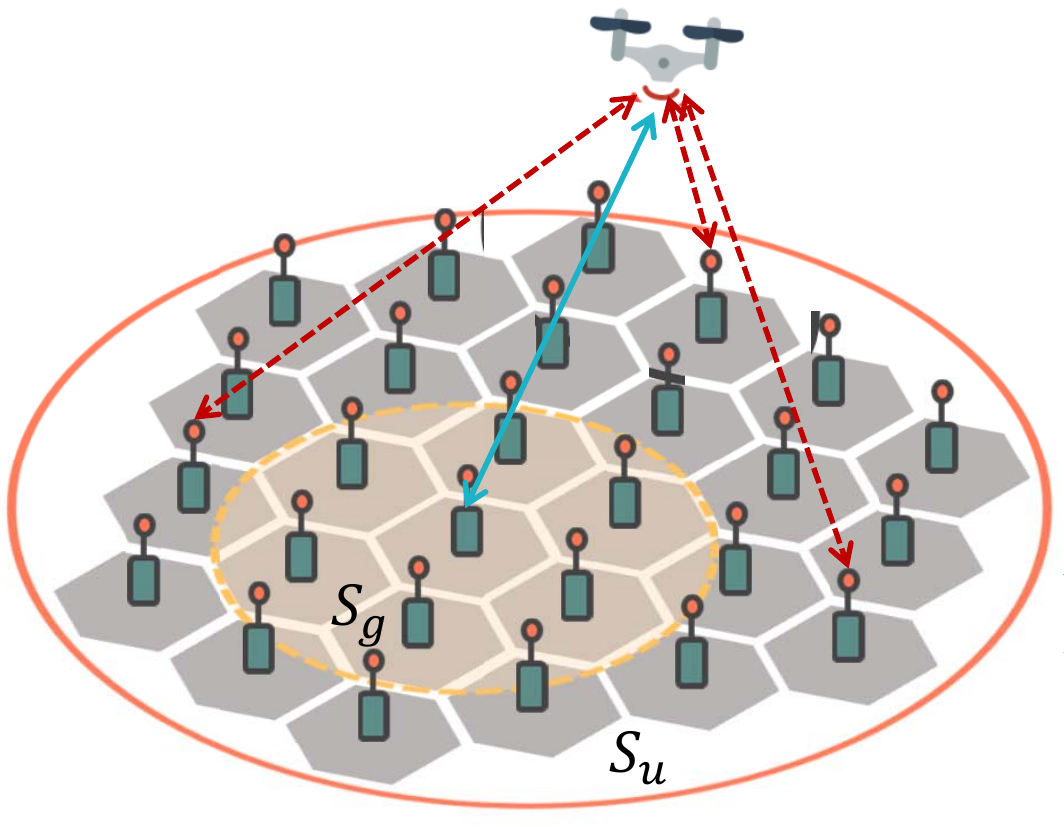}
	\caption{\label{fig:UAVIC} An aerial BS interferes with a much higher number of cells than a ground BS. Here, $S_g$ is the area interfered with a ground BS and $S_u$ is the area interfered with a UAV BS. }
\end{figure}

 Such an approach (treating interference as noise) results in poor transmission rates unless the interference power is very small, which is not the case in today's networks. This approach will cause poorer performance in 3D networks as co-channel interference is much stronger and much more prevalent in 3D cellular networks.  
As an example, while the average number of neighbors for a user at a height of 1.5$\,m$ is 5, this number increases to 17 at the height of 120$\,m$ \cite{amorim2017radio}, as 
depicted in Fig.~\ref{fig:UAVIC}. 

To tackle this growing issue, this article advocates the use of \textit{deep reinforcement learning} (DRL)-based interference mitigation as an alternative solution. DRL is a great tool to approach this problem  for several reasons: it is inherently useful for decision-making in complex and dynamic environments like a multi-cell network; it can outperform traditional approaches when (near)-optimal solution is elusive; and it can learn from traditional approaches (e.g., CoMP and interference alignment) while potentially performing better than them with less stringent CSI requirements.

 Despite its significant potential, 
DRL-based co-channel interference management in cellular networks has been explored in limited cases. 
 This motivates us to summarize the main challenges and potential solutions
for developing novel DRL-based interference management techniques in this article to inspire future
research in this field. We particularly emphasize practical assumptions in terms of the availability of CSI and the scale of the network in our solutions. An ultimate goal is to develop solutions that are independent of the CSI of interfering signals and compatible with 3GPP signal/interference
power estimation methods, particularly those measured in  Long-Term Evolution (LTE) and New Radio (NR) networks  \cite{RSSI}.

We discuss solutions to  mitigate interference in 3D cellular networks in generic settings in terms of the number of cells, the number of antennas, and the frequency of operation (sub-6 GHz and millimeter wave (mmWave) bands). We also argue that the solutions can be scalable with sub-linear complexity by developing novel multi-agent learning and actor-critic architectures. 
Further, numerical results are provided to evaluate and validate the effectiveness of DRL-based interference management  without requiring the CSI of the interfering links. 
We also discuss the 3GPP standardization aspects of embracing DRL for interference
management in UAV-based 3D networks.
Finally, we highlight several future
research directions and conclude the paper.


\section{3D Networks: A Primer}

\subsection{UAV Networks: Common Use Cases}

In general, there are two ways that UAVs can be exploited in cellular networks: aerial users and aerial infrastructures.

\begin{figure*} 
	\centering
	\includegraphics[width=.85\textwidth]{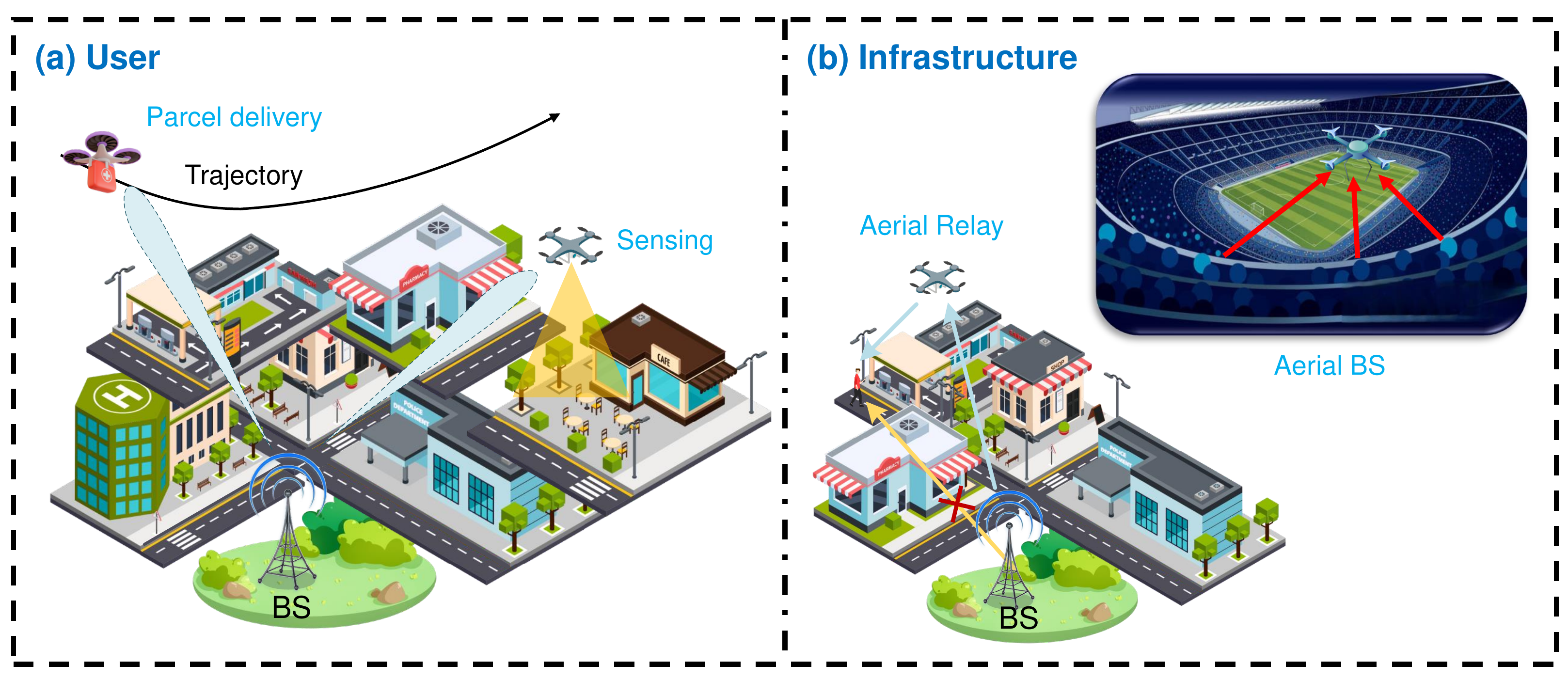}
	\caption{\label{fig:use} Common use cases for UAVs: aerial users and infrastructures.}
\end{figure*}

\textbf{User:} As shown in Fig. \ref{fig:use} (a), one common use case of UAVs is to deliver parcels. In this case, UAV is a user equipment (UE).  To guarantee  the safe operation of UAVs, cellular networks with wide coverage can control the trajectory of these UAVs. The UAVs could also be equipped with various sensors to execute sensing tasks due to their on-demand deployment and larger service coverage compared with the conventional fixed sensor nodes. As the altitudes of UAVs are typically much higher than 
the base station (BS)'s antenna height, they are identified as a new type of user that requires 3D coverage, as opposed to the conventional 2D ground coverage. 

\textbf{Infrastructure:} UAVs can also be used as new types of aerial infrastructures, including BSs and relays, to cover hard-to-reach areas, as illustrated in Fig. \ref{fig:use} (b). For example, in hotspots or disaster areas where the terrestrial infrastructure is destroyed or users are underserved, UAVs can serve as temporary access points for emergency communication or data offload to further improve the performance of terrestrial wireless communications.


\subsection{Channel Model and Propagation Environment}

Depending on the transmission modes, the channel model of a UAV can be categorized into two types, i.e., air-to-ground  and air-to-air channels. Since the communications between UAVs typically occur in clear airspace, the air-to-air channel can be characterized by the free-space path loss model.
The air-to-ground channels significantly differ from those used in terrestrial communications. Their characteristics highly depend on the altitudes and elevation angles of the UAVs, and any movement caused by the UAVs will impact the channel.  Since an air-to-ground channel can also be occasionally blocked by obstacles such as terrain, buildings, or the airframe itself, a larger elevation angle leads to a lower path loss as the line-of-sight components will be more likely to dominate. To model these characteristics, statistical path loss models have been widely used for the air-to-ground channels, where line-of-sight and non-line-of-sight components are considered to occur with different probabilities \cite{zhang2020unmanned}. It should be noted that the reciprocity of air-to-ground channels still holds although the ground BS will optimize its tilt angle to maximize the performance of ground users. This is because the change of tilt angle only influences the antenna gains in an angle span.

\subsection{Inter-cell Interference in UAV Networks}
Several techniques have been proposed in the literature to alleviate inter-cell interference in UAV-based networks. These include:  

\subsubsection{Time/Frequency orthogonalization}
The time/frequency resource is divided into resource blocks and each resource block is assigned to a UAV within a cell at the same time. Although UAVs associated with different BSs can share the same resource block, the reuse factor needs careful optimization to further reduce the interference as inter-cell interference is more significant in UAV networks.  

\subsubsection{Beam orthogonalization}
Spatial techniques are used in this case. Specifically, UAVs equipped with multiple antennas can generate directional beams toward the target receiver and alleviate the interference to neighboring cells by suppressing the side lobes. Even with a single antenna, a UAV swarm can also form a virtual multiple-input multiple-output (MIMO) for beamforming \cite{hanna2021uav}.

\subsubsection{Path Design}
Due to the mobility of UAVs,  the paths of UAVs can be designed to make the distance between the UAVs sharing the same spectrum as larger as possible to reduce mutual interference \cite{Challita2019Deep}. 

\subsubsection{Cooperation Techniques}
Neighboring BSs can cooperate to decode the signals transmitted from UAVs. For example, each BS will decode the signals from its associated UAVs and exchange the decoded results with neighboring BSs, which is the inter-cell interference to be canceled \cite{liu2019comp}. 

Existing inter-cell interference management methods require accurate CSI for both desired and interfering links, which is expensive to obtain in practice.
This issue motivates exploring new tools for treating the multi-cell interference problem.

\section{DRL for Interference Management: Why and How?}
\label{sec:DRL}

 

\subsection{Why Deep Reinforcement Learning?}

Reinforcement learning  is the science of decision-making \cite{sutton2018reinforcement}. In reinforcement learning an \textit{agent}
 learns to
interact with an \textit{environment} by taking a sequence of \textit{actions}  to maximize cumulative \textit{reward}. 
The ultimate goal is to maximize the \textit{utility} which is an estimation of the long-term reward that the agent is expected to receive. For this, we will need to learn a \textit{policy} function $\pi$ that maps states to actions.
The \textit{value} of a state ${\bm s}$ is quantified by $V({\bm s})$, which is the utility we expect to get if we are at state ${\bm s}$ and play optimally. A related more frequently seen quantity is the state-action value $Q_\pi({\bm s},{\bm a})$, which is the expected utility starting at state ${\bm s}$, taking action ${\bm a}$, and  playing optimally thereafter.
Then, we learn  $Q$-values and extract actions from them. In $Q$-learning,  $Q$-values are iteratively computed according to 
the Bellman equation \cite{sutton2018reinforcement}.
When state-action space is large, deep \textit{Q}-learning is preferred to tabular \textit{Q}-learning because it can  reduce the computational complexity by avoiding an exhaustive search in the action space \cite{sutton2018reinforcement}.
 In DRL, deep neural networks are used to approximate the value, the policy, or both.  DRL increases the learning capacity of reinforcement learning and enlarges its scope by removing the need for expert feature engineering to train the algorithm.  
 
 DRL has great potential for solving problems whose optimal solution is unknown or requires poor approximation.  Cellular interference management is a notable example of such problems. Conventional interference management solutions such as  interference alignment, CoMP, \textit{signal-to-interference-plus-noise ratio} (SINR)  maximization are sub-optimal and highly depend on the accuracy/availability of the interference CSI. 
  By properly defining the agent, environment, state, action, and reward, not only can DRL learn how to implement traditional interference  management algorithms, but it can also surpass them. As we will see in this article, using DRL   

 \begin{itemize}
     \item multiple traditional algorithms can be combined in one learning model via multi-objective learning
     \item the learning process can be completed online and without  a stringent need on CSI of interference
     \item learning is inherently robust to the dynamic environment of 3D multi-cell networks, e.g., it works if a line-of-sight channel becomes non-line-of-sight  or vice versa, whereas interference alignment would fail.  
 \end{itemize}
Through interaction with the environment (multi-cell network), the agent learns how to make better decisions to maximize a cumulative reward via taking a sequence of actions.  The interference management is handled by an agent, which can be either centralized or distributed based on the specific DRL algorithm being used.
In a centralized DRL, the agent would be a central node such as a BS controller (BSC) in 2G, the radio network controller (RNC) in 3G, or a central radio resource management controller  in 4G/5G.  Alternatively, the agent could also be one of the BSs itself. If UAVs serve as BSs, the agent may be referred to as the UAV controller. This controller can either be one of the UAVs or a separate entity that communicates with the UAVs to issue actions. 
As the agent learns to maximize its explicit objective (i.e., the cumulative reward), it can maximize other implicit optimizations as well, such as network sum-rate or spectral efficiency.
 This  power of DRL makes it very competitive for interference management. In a distributed DRL, the BSs together make the agent.

\begin{figure*} 
\centering
 	\includegraphics[width=.8\textwidth]{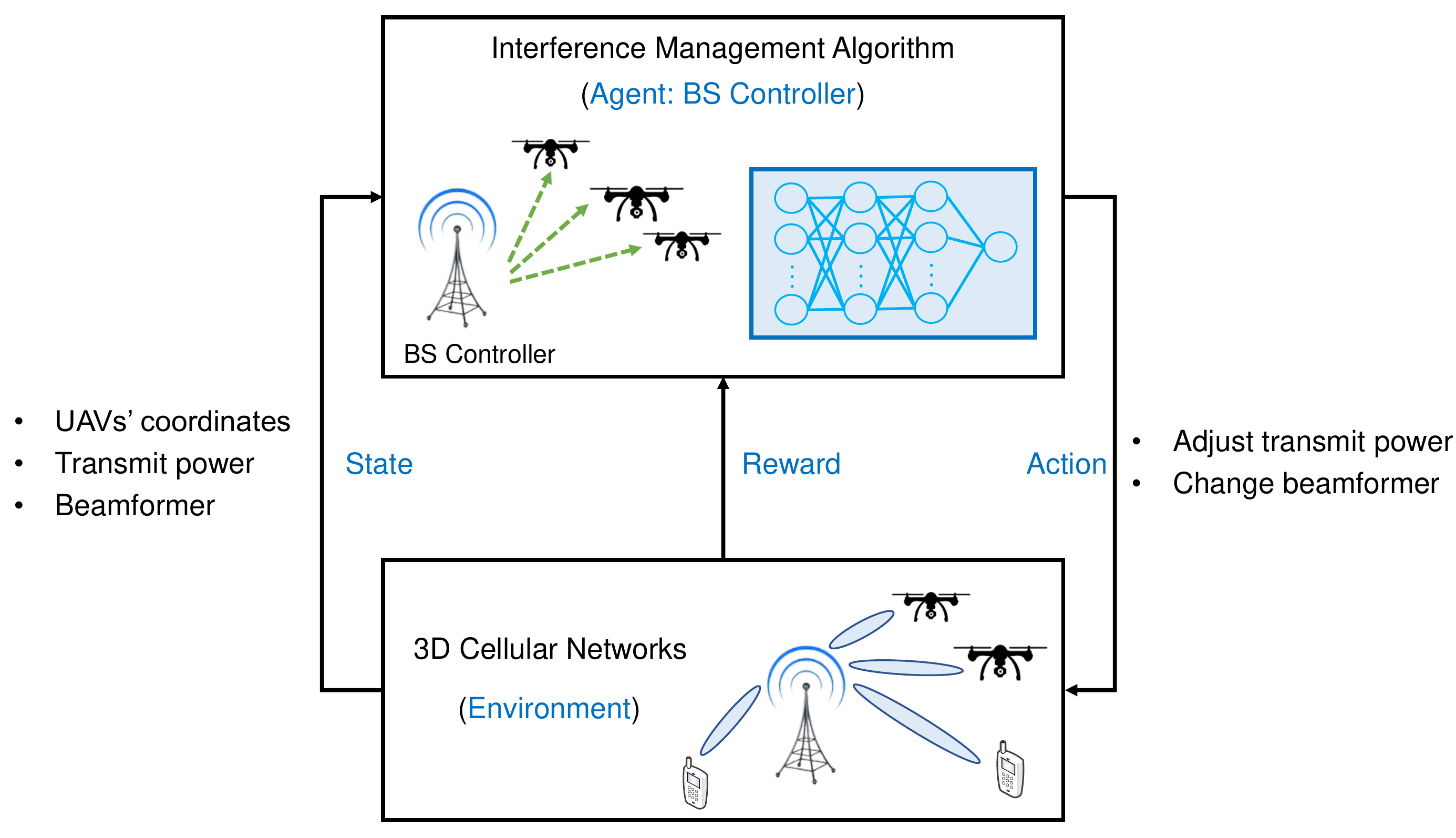}
\caption{Agent-environment interaction in reinforcement learning based multi-cell network. The agent is a BS controller and depending on the UAVs' use case it may control aerial BSs (UAVs) or ground BSs.}
		\label{figDRL} 
\end{figure*} 



DRL can be leveraged to develop novel interference mitigation algorithms and improve spectral efficiency with real-world assumptions and constraints, in terms of the availability of CSI, range of signal-to-noise ratio (SNR), network scale, and stochastic nature of the environment.  There are two important questions that we seek to answer here: 
\textbf{(i)} How can we mitigate the interference with local or no CSI? 
\textbf{(ii)} How to make the algorithms work with the large network scale and high-dimensional state and action spaces?    

 These questions can be explored in various settings in terms of the operating frequency band (sub-6 GHz/mmWave), propagation environment (line-of-sight, non-line-of-sight, and mixed), the level to which
CSI is available, number of antennas at each node, and network scale. Network design and training will be different in each case. For example, mmWave/Terahertz MIMO systems rely on predefined coodbooks whereas sub-6~GHz systems use digital beamforming. The former implies a \textit{discrete} domain whereas the latter is \textit{continuous}. DRL algorithms for the two cases have to be designed differently.

\subsection{An Illustrative  Example }
\label{subsec:example}
Consider a downlink cellular network where $L$ users simultaneously communicate over a wireless channel in $L$ cells.
Let each BS have $M$ antennas and  serve single-antenna users. 
To make the example less involved, let us assume mmWave transmission where analog-only beamforming is common. Then, we will have a discrete 
set for beamforming vectors $\mathbf{w}_{\ell} $.  
Let $\mathcal {W}$  be the  \textit{beamforming codebook} adopted by the BSs. 
Further, let $P_{\ell}$ represent the transmit power of  BS~$\ell$, and assume that the set of BS powers, denoted by  $\bf \mathcal {P}$, is discrete.


The goal is to find $P_{\ell} \in \mathcal {P}$ and $\mathbf{w}_{\ell} \in \mathcal {W}$ so that the network sum-rate, which is a common measure of spectral efficiency in cellular networks, is maximized.
The network sum-rate is defined as ${\sum_{\ell = 1 }^L \log_2(1+\gamma_\ell)}$ where $\gamma_\ell$ is the SINR at the UE located at cell~$\ell$.  The SINR evaluation requires the CSI for both the desired and interference signals. 
Hence, at each BS we need to know global CSI, i.e., $\mathbf{h}_{\ell,j}$  for all $\ell$ and $j$. 
 This is not, however, practical since CSI overhead will consume a big portion of the bandwidth particularly when $L$ is large.    
In addition, the above optimization problem is nonconvex and is hard to solve, and the exhaustive search and other traditional optimization methods need global CSI knowledge.

Next, we develop an alternative solution based on DRL.  The proposed method can work based on limited knowledge of CSI (e.g., with only the serving cell channel $\mathbf{h}_{\ell,\ell}$, local CSI, or even no explicit CSI)  
which is a big advantage compared to the global CSI.
As shown in  Fig.~\ref{figDRL},  the environment is a multi-cell network; the agent is a BS controller which implements an interference management algorithm (e.g., CoMP, SINR  maximization, or others), and the reward could vary depending on the availability of the CSI but is supposed to help increase the network sum-rate. 
To describe the potentials and challenges let us define the state, action, and reward.   
	\begin{itemize}				
	\item The \textit{state} ${\bm s}_t \in \mathcal{S}$ is a representation of the environment that describes the current situation. It is what an agent observes at time $t$. Let ${\bm s}_t$ be the collection of UAVs' coordinates, BSs' powers and beamforming vectors.
	Even if we assume all the five elements $x_\ell$, $y_\ell$,  $z_\ell$, $P_\ell$ and $\mathbf{w}_{\ell}$ are  discrete and each can take only 10 distinct values, 
	the number of states will be $|\mathcal{S}| = 10^{5L}$, which is extremely high.

	\item The action ${\bm a}_t \in \mathcal{A}$ is the move taken by the agent within the environment at time step $t$. The action ${\bm a}_t$ will advance the state ${\bm s}_t$ to  ${\bm s}_{t+1}$.  
	In this example, actions are to change the power and beamforming vector of each BS.
	Let action ${{\bm a}_t }$ be a binary vector $
{{\bm a}_t}  = \left [a_1,\dots, a_\ell, \dots, a_{2L} \right]
$ where each element being either \lq0\rq or \lq1\rq. More specifically, for any $\ell \in \{1, \dots, L\}$, we have
	\begin{itemize}
		\item 
		$a_\ell = 0$: decrease the transmit power of  BS~$\ell$ by 1dB,
		\item $a_\ell = 1$: increase the transmit power of  BS~$\ell$ by 1dB,
		\item $a_{L+\ell} = 0$: step down the beamforming  index of BS~$\ell$, and
		\item $a_{L+\ell} = 1$: step up the beamforming  index of BS~$\ell$.
	\end{itemize}
	The cardinality of  actions is $\mathcal{|A|} = 2^{2L}$. Clearly, by taking action ${\bm a}_t$, the agent is changing the  beamforming vectors and transmit powers for the serving and interfering BSs. Thus, this is a collaborative interference management scheme via coordinated power and beamforming design.

		\item  The \textit{reward} is a mechanism telling the agent the consequence of its actions. The agent’s goal is to take actions that maximize
		an estimation of the long-term reward it is expected to receive (simply, the total cumulative  reward).

\end{itemize}

Defining  the reward function is a crucial step in determining the type of CSI (global, local, or no CSI) needed.  
Since our goal is to maximize the sum-rate of the multi-cell network, the reward  
would ideally be based on the sum-rate.  
With this, the agent must evaluate the SINR received by the UEs at all cells which requires global knowledge of CSI.
At each time step, if $\gamma_\ell > \gamma_{\min}$ for all users, then the reward  would be $\sum_{\ell=1}^{L}\gamma_\ell$; otherwise, we give a penalty to ensure the quality of service for all users.

We propose using rewards that can mitigate inter-cell interference without requiring the CSI of the interference links. One example of such rewards is to use SNR instead of SINR in reward calculations  which requires serving CSI only. Even with this simplified method, the DRL algorithm can effectively infer the severity of interference from the users' coordinates and BSs' power to adapt its actions based on it. For instance, it can deduce that a user located at the cell edge will cause more interference than the one located at the cell center. 
In Section~\ref{sec:reward}, we provide additional examples of rewards that do not require CSI of interference links.

\begin{figure} 	
	\centering
\includegraphics[width=.48\columnwidth] 
{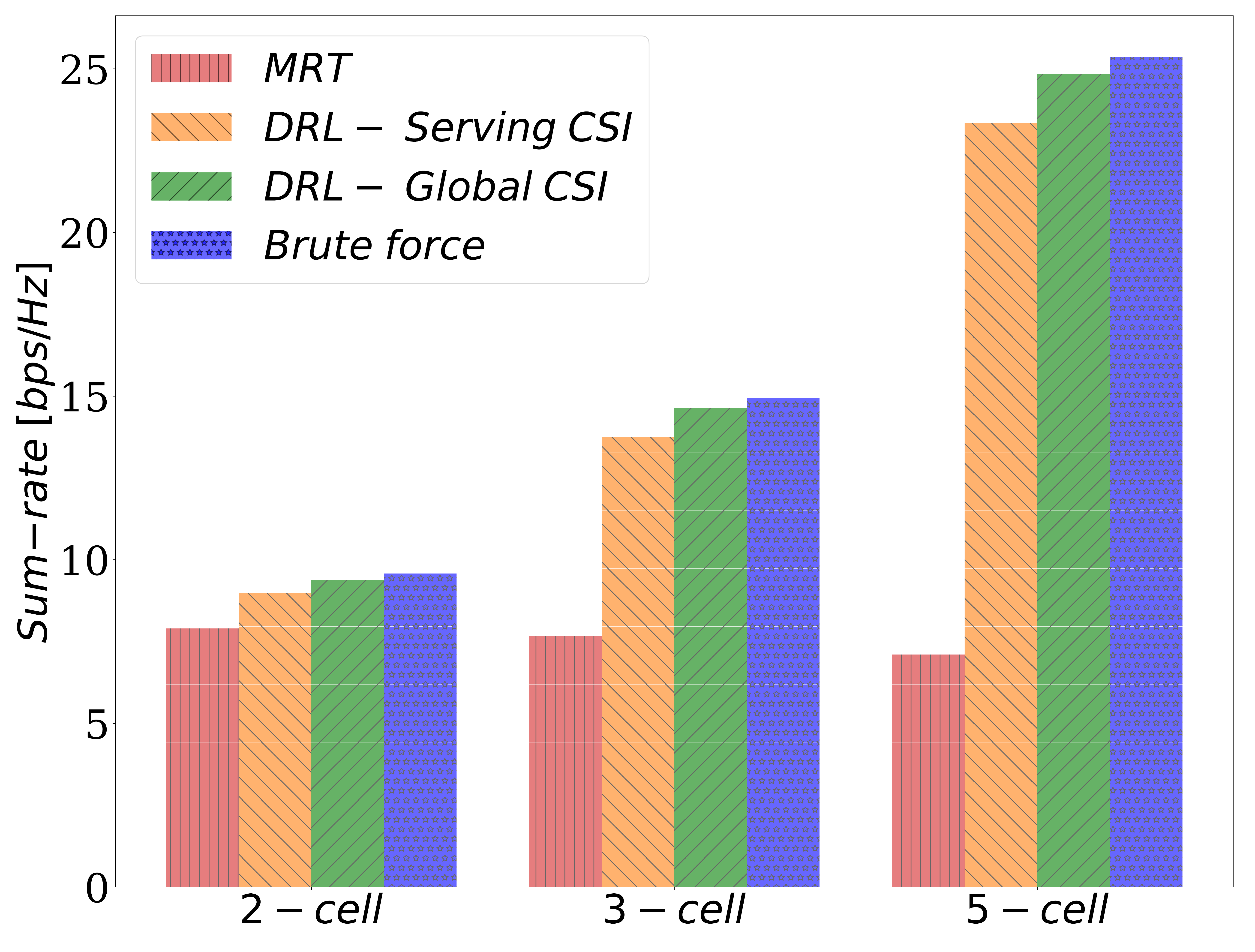}
\caption{\label{fig:DRL_GC} Network sum-rate using  DRL-based approach with serving and global CSI, MRT beamforming, and brute force search with global CSI for the different number of cells.}
\end{figure}

\begin{figure}[t] 
	\centering
  \includegraphics[width=.5\columnwidth]
  {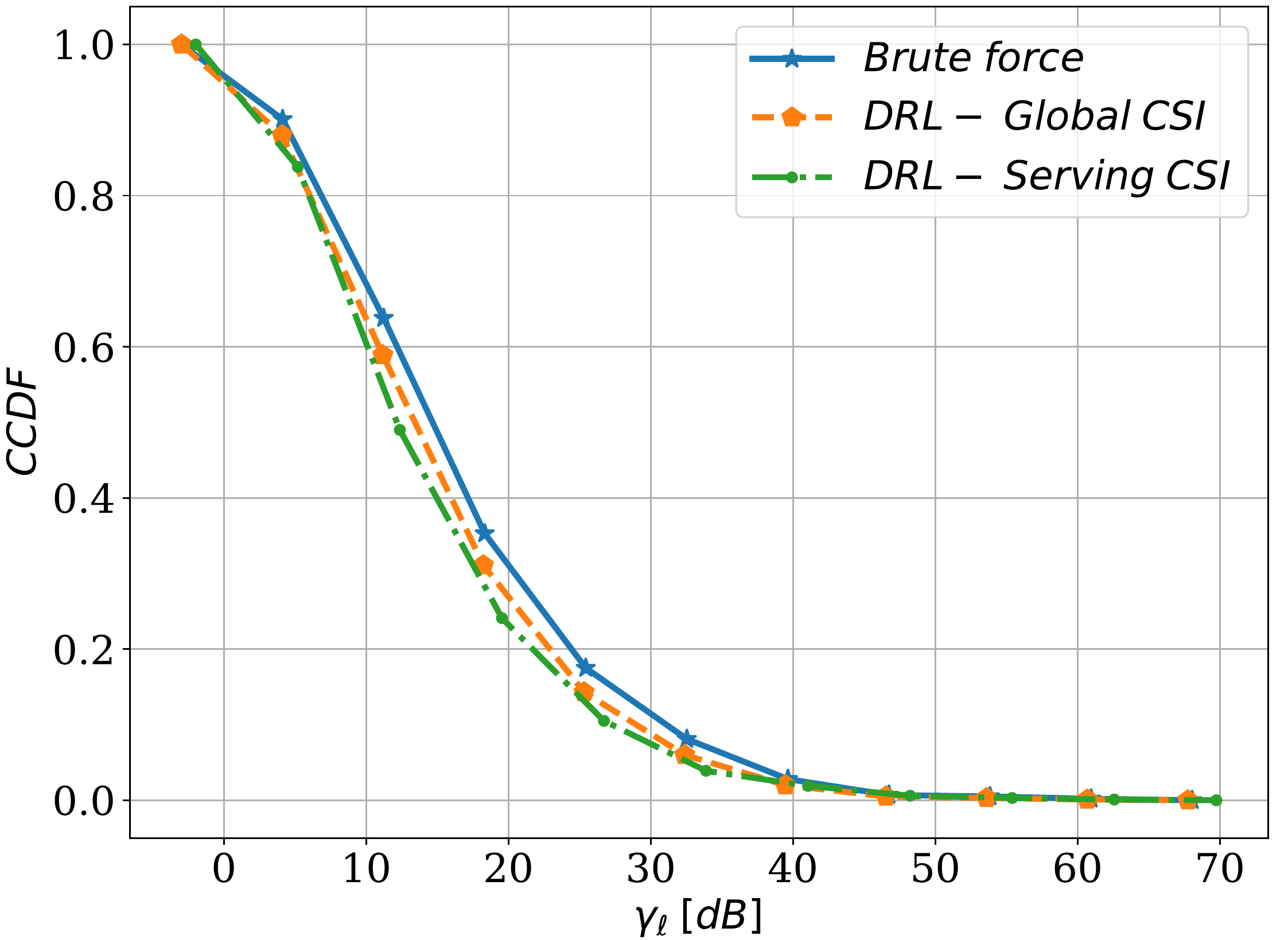}
\caption{\label{fig:SINRlocalCSI} CCDF of coverage of DRL with  serving and global CSI as well brute force method for $L=5$.}
\end{figure}

Next, we show the effectiveness of the proposed method via simulations. Spectral efficiency and overall network coverage are the performance metrics we use for the evaluation. The former is measured by the achievable sum-rate whereas the latter is evaluated by the \textit{complementary cumulative distribution function (CCDF)} of the SINR.  The results are plotted in Fig.~\ref{fig:DRL_GC} and Fig.~\ref{fig:SINRlocalCSI}, where DRL-based solutions are compared with maximum ratio transmission (MRT) beamformer \cite{bjornson2010cooperative} and brute force search.  
The proposed DRL algorithm learns to mitigate interference with serving CSI almost as effectively as that with global CSI. The brute force method also uses global CSI knowledge to find the best beam and power at each BS. MRT is oblivious to interference since it replaces the $L$-cell interference problem with $L$ single-cell problems. As a result, its spectral efficiency does not scale with $L$. 
In this experiment, 
users are uniformly distributed within the cells, and the probability of LoS for the channels is 0.8.   We set $\gamma_{\min} =-3$dB. Any channels that do not meet this threshold will not be scheduled and will be dropped. To be fair, the same channels that fail to meet this threshold are also excluded from MRT calculations. However, when $L$ increases, there is a reduction in MRT gain due to the possibility of some good channels being dropped because of high interference.  If users are pushed toward the cell edge,  interference will be higher and sum-rates will reduce for all methods except for MRT, as it avoids interference using time division multiplexing.  

\subsection{Other Types of  Rewards}
\label{sec:reward}
The above example illustrates the basic idea of DRL-based interference mitigation which does not require the CSI of the interference links and relies on serving CSI only. 
 One can use more competitive rewards such as:

   \subsubsection{No-CSI reward} In this case, SINR will be measured using local power measurements without explicitly requiring CSI. To do this, when the serving BS is not transmitting, at each cell the UE will receive and measure  interference plus noise ($\rm I+N$) level. Next, when serving BS is transmitting, the UE can measure  signal plus interference plus noise ($\rm S+I+N$\rm ). Subtracting these two measurements, the UE can find signal power ($\rm S$) and evaluate $\rm SINR= S/(I+N)$. 


    \subsubsection{3GPP compatible rewards} A 3GPP compatible reward may use signal
power and interference estimation measurements like \textit{received signal strength indicator} (RSSI) 
and \textit{reference signal received quality} (RSRQ) \cite{RSSI}. These are key signal-level measures of LTE/NR networks. 
This way, similar to the no CSI case disused earlier, we do not need to know any CSI  explicitly.

\subsubsection{Compound rewards}
 Besides the  above signal level and quality measures which are common in LTE/NR networks, SINR is also measured in NR networks \cite{RSSI}. RSRQ and SINR measures are related to the interference and can be used along with RSSI to define compound rewards which are based on multiple measurements rather than a single one. 

\subsubsection{Multi-objective rewards} 
Here, the agent can have multiple objectives  each with its own rewards. For example, we  may consider interference mitigation and UAV trajectory planning as two objectives.

\section{Open Challenges}

The DRL-based algorithm of Section~\ref{sec:DRL}
 works for discrete state/action space  
where the number of  beamforming vectors is limited and the goal is to show the potential of DRL with local CSI. In this section, we point out some fruitful avenues for extending and generalizing the above example.

\subsection{Scalability}
  The state and action spaces could be \textit{discrete} or \textit{continuous}. 
Even in the discrete case, the state space is typically very large and \textit{exponentially} increases with the number of cells. 
In a tabular $Q$-learning, the state-action function $Q_\pi(\mathbf{s}, {\bm a})$
is represented by a table of size $\mathbb{R}^{|\mathcal{S}| \times |\mathcal{A}|}$. 
In our example in the previous section, we had  $|\mathcal{S}| = 10^{5L}$ and $|\mathcal{A}| = 2^{2L}$ which result in $10^{90}$ possible states and $2^{36}$ possible actions for  $L=18$.     Then, even for this  simplified example, the tabular $Q$-learning is impractical for our desired problem. 

To address this difficulty, function approximation can be used to estimate the value of states or actions. 
Deep Q-network (DQN) uses deep neural networks as a function approximation.
 Function approximation using DQN makes learning more efficient and allows us to reason about previously unseen actions. However, DQN has stability issues and its complexity grows \textit{linearly} with  $|\mathcal{A}|$. This renders DQN intractable when the number of actions is significant, which is the case in 3D multi-cell networks. 
 To overcome this, we propose Wolpertinger-based learning. Wolpertinger architecture  is an effective way of reasoning when the action space is huge \cite{dulac2015deep}. It generalizes over action space with a \textit{sub-linear} complexity.  
 This method generates actions that may not be a valid action. 
 It then uses the $k$-nearest neighbor clustering to map from a continuous action space to a discrete set.

\subsection{Continuous State-Action}
Discrete spaces are plausible when we choose beamforming vectors from a discrete set, which is common in mmWave analog-only beamforming \cite{hur2013millimeter}. With digital beamforming, which is common in sub-6GHz communication, 
beamforming vectors take values from a continuous domain.
One approach is to discretize the spaces. This will however make learning difficult due to noise and delayed reinforcements. An alternative is learning from a continuous space.
Value-based DRL is not suitable to model continuous action space (e.g., digital beamforming). A policy-based DRL may be applied alternatively.  
Unlike value-based methods, policy-based methods remain stable under function approximation, but they suffer from sample inefficiency. 
An \textit{actor-critic} algorithm \cite{lillicrap2015continuous} is a powerful approach that combines the two methods. In such an algorithm,  the policy (actor) and value (critic) functions are parameterized to enable the effective use of training data with stable convergence.

\subsection{Distributed Learning}
A single agent implies a backhaul connection for the communication of the DRL agent and the BSs. While the overhead of this communication may not be large (it is 2 bits/BS in our example in our earlier example), distributed learning is preferred to avoid the exchange of this information. Multi-agent DRL can accomplish this.

In multi-agent DRL, multiple agents interact within a common environment to learn a policy for each
agent such that all agents together achieve the desired goal of the system. 
 The main advantage of multi-agent DRL in a multi-cell network is that each BS (agent) learns to shape the signal of its user independently. 
If we consider each BS as an independent learner, the other agents’ actions would be treated as part of the environment. 
Cooperative multi-agent algorithms are the other extreme. In this approach, the agents learn to share their learning. This will, however, increase the communication overhead.
 We propose using sequential learning of the agents without observing the actions of other agents and without complicated communication. We can order the neighboring cells by the severity of interference they receive 
 (e.g., based on the distance of the user to the neighboring BSs or RSRQ) 
 and train them sequentially, after training the serving agent. 
 The serving agent can allocate any beam whereas the neighboring agents choose the beam that causes the least interference. 
Such algorithms also decrease the complexity from $| \mathcal{A}|^L$ to $L| \mathcal{A}|$, where  $L$ is the number of cells and
 $| \mathcal{A}|$ is the number of actions in each cell.


\subsection{Mobility and Doppler Shift}

In our case study, the moving speed is assumed to be low so that we can approximate the UAV as static in each time slot. However, sometimes, for instance, if the UAV is delivering a parcel, the moving speed of the UAV could be high. In such a case, the high mobility might introduce a Doppler shift, thus introducing inter-carrier interference. 

To address this issue, a solution is to include the parameters of the mobility in the input of the deep neural networks, e.g., the moving direction and speed. Therefore, the structure of the neural networks also needs to be well designed to weigh the impact of mobility/Doppler shift properly on the inter-carrier interference. We can also include the moving of the UAV in the action space so that the DRL algorithm can tell what the optimal path is to have minimum interference. 

 \subsection{Model-based DRL}
 
We have considered model-free reinforcement learning algorithms so far since it is hard to get a ground-truth model of the dynamic environment of multi-cell networks. 
In model-free DRL, the algorithm estimates the optimal policy without using or estimating the dynamics (transition and reward functions) of the environment. 
 On the other hand, a model-based DRL uses a known or learned model (e.g.,  $p({\bm s}_{t+1}|{\bm s}_{t},{\bm a}_{t})$ probability of transiting to the next state) when learning to approximate a global value or policy function. Although finding or learning a decent model
for the multi-cell interference mitigation task is demanding, model-based DRL has a big advantage. It has a much higher sample efficiency, and thus, is far less complex. To train an 18-cell, 3D network with a policy gradient method may take several days while model-based DRL may take less than an hour.

\section{3GPP Standardization Aspects}
\label{sec:3GPP}

The interplay between DRL and interference management in UAV-based 3D networks and the standards work on UAV communication in 3GPP is an interesting topic of practical relevance. On the one hand, DRL algorithms for interference management in UAV-based 3D networks can leverage the latest standards features developed by 3GPP. On the other hand, the 3GPP standards work on UAV communication can evolve towards embracing DRL for interference management in UAV-based 3D networks.

\subsubsection{Existing 3GPP work on interference management for LTE- and NR-connected UAVs}

In Release 15, 3GPP conducted a study item on enhanced LTE support for aerial vehicles, assessing the performance of utilizing LTE networks to provide UAV connectivity. The outcome of the study highlighted that interference issues exist in both uplink and downlink when providing cellular connectivity to UAVs, particularly for dense UAV scenarios. The study identified a set of solutions for interference detection and interference mitigation to address the interference problems. In a follow-up work item, 3GPP introduced specification enhancements to improve the performance of LTE-connected UAVs.
Compared to LTE, 5G NR has significantly improved capabilities and can provide efficient UAV connectivity in more diverse scenarios. To further improve the 5G networks' capabilities for UAV communication, 3GPP is conducting a work item in Release 18 to introduce UAV-related enhancements in NR specifications. 
In Release 15, the study item on enhanced LTE support for aerial vehicles identified that using directional antennas at an aerial UE can help mitigate the interference problems in both uplink and downlink. However, it was considered at that time that the use of directional antennas at the aerial UE was an implementation issue. Thus, there was no corresponding specification enhancement introduced in LTE. However, 3GPP revisits this topic in the Release-18 NR UAV work item and studies UE capability signaling to indicate UAV beamforming capabilities and, if necessary, radio resource control signaling for UAV UE with a directional antenna.

\subsubsection{Potential future 3GPP work on DRL-based interference management for connected UAVs}

Though some DRL-based interference management solutions (e.g., the solution presented in Section~\ref{sec:DRL}) may be purely based on proprietary implementations by exploiting the existing features in the standards, others may benefit from further specification enhancements. Along this line of work in 3GPP, embracing DRL for interference management in UAV-based 3D networks can be an interesting evolution direction in 3GPP. It is worth noticing that 3GPP Release 18 studies artificial intelligence (AI)/machine learning (ML) for the 5G NR air interface. The study investigates the 3GPP framework of AI/ML for air interface under three selected use cases, including CSI feedback, beam management, and positioning. The selected use cases represent generic functionalities, and the corresponding potential enhancements can be leveraged to improve interference management for UAV communication. The current selection of the three use cases targets formulating a framework to apply AI/ML for the NR air interface. It is anticipated that 3GPP would investigate more use cases to apply AI/ML for the air interface. Such future use cases may include specific features dedicated to interference management for UAV communication. 

\section{Conclusions}
In this article, we have illustrated a concrete example on the use of DRL for interference mitigation  without requiring the CSI of interfering signals and indicated how this solution can be extended in various other settings. 
Overall, we have shown that this framework can be used to explore important questions surrounding interference management in UAV-based 3D networks, such as making the algorithms scalable and having the spectral efficiency grow with the number of cells.  In addition, we have discussed using 3GPP-based reward with no explicit CSI 
and having multi-objective learning by combining those rewards and path planning to avoid interference besides SINR maximization.

\vspace{-1cm}
\begin{IEEEbiographynophoto}
{\bf Mojtaba Vaezi} (mvaezi@villanova.edu) is an Assistant Professor in the ECE Department at Villanova University.  
 His interests include machine learning for wireless networks. He was a visiting fellow at Princeton University in 2022.  He was the recipient of the 2020 IEEE Communications Society Fred W. Ellersick Prize.
\end{IEEEbiographynophoto}
\vspace{-1.5cm}
\begin{IEEEbiographynophoto}
{\bf Xingqin Lin} (xingqinl@nvidia.com) is a Senior Standards Engineer at NVIDIA, leading 3GPP standardization and conducting research at the intersection of 5G/6G and AI. Before joining NVIDIA, he was a Master Researcher and a member of NextGen Advisory Board at Ericsson.
\end{IEEEbiographynophoto}
\vspace{-1.5cm}
\begin{IEEEbiographynophoto}
{\bf Hongliang Zhang} (hongliang.zhang@pku.edu.cn) is an assistant professor in the School of Electronics at Peking University. He was the recipient of the 2021 IEEE Comsoc Heinrich Hertz Award.
\end{IEEEbiographynophoto}
\vspace{-1.5cm}
\begin{IEEEbiographynophoto}
{\bf Walid Saad} (walids@vt.edu)  is a
professor in the Department of Electrical and Computer Engineering at Virginia Tech. He is the co-author of ten conference best paper awards. His research interests include wireless networks, machine learning, game theory, cybersecurity, unmanned
aerial vehicles, and cyber-physical systems.
\end{IEEEbiographynophoto}
\vspace{-1.5cm}
\begin{IEEEbiographynophoto}
{\bf H. Vincent Poor} (poor@princeton.edu) is the Michael Henry Strater University Professor at Princeton University, where his interests include wireless networks and related fields. A member of the U.S. National Academies of Engineering and Sciences, he received the IEEE Alexander Graham Bell Medal in 2017.
\end{IEEEbiographynophoto}


\begin{thebibliography}{10}
	
	\bibitem{amorim2017radio}
	{R. Amorim \emph{et al.}}, ``{Radio channel modeling for UAV communication over
		cellular networks},'' {\em IEEE Wireless Commun. Lett.}, vol.~6, no.~4,
	pp.~514--517, 2017.
	
	\bibitem{3GPP}
	{3GPP}, ``{Technical specification group radio access network: Study on
		enhanced LTE support for aerial vehicles},'' {\em 3GPP 36.777 V15.0.0}, 2018.
	
	\bibitem{antennagain}
	{X. Lin \emph{et al.}}, ``{The sky is not the limit: LTE for unmanned aerial
		vehicles},'' {\em IEEE Commun. Mag.}, vol.~56, no.~4, pp.~204--210, 2018.
	
	\bibitem{Challita2019Deep}
	U.~Challita, W.~Saad, and C.~Bettstetter, ``{Interference management for
		cellular-connected UAVs: A deep reinforcement learning approach},'' {\em IEEE
		Trans. Wireless Commun.}, vol.~18, no.~4, pp.~2125--2140, 2019.
	
	\bibitem{jafar2011interference}
	S.~A. Jafar, {\em Interference Alignment: A New Look at Signal Dimensions in a
		Communication Network}.
	\newblock Now Publishers Inc, 2011.
	
	\bibitem{irmer2011coordinated}
	{R. Irmer \emph{et al.}}, ``Coordinated multipoint: Concepts, performance, and
	field trial results,'' {\em IEEE Commun. Mag.}, vol.~49, no.~2, pp.~102--111,
	2011.
	
	\bibitem{RSSI}
	``{NR physical layer measurements (Release 17)},'' {\em 3GPP TS 38.215
		V17.1.0}, [online] Accessed on June 2022.
	\newblock
	\url{https://www.3gpp.org/ftp/Specs/archive/38_series/38.215/38215-h10.zip}.
	
	\bibitem{zhang2020unmanned}
	H.~Zhang, L.~Song, and Z.~Han, {\em Unmanned Aerial Vehicle Applications over
		Cellular Networks for 5G and Beyond}.
	\newblock Springer, 2020.
	
	\bibitem{hanna2021uav}
	S.~Hanna, E.~Krijestorac, and D.~Cabric, ``{UAV swarm position optimization for
		high capacity MIMO backhaul},'' {\em IEEE J. Sel. Areas Commun.}, vol.~39,
	no.~10, pp.~3006--3021, 2021.
	
	\bibitem{liu2019comp}
	L.~Liu, S.~Zhang, and R.~Zhang, ``{CoMP} in the sky: {UAV} placement and
	movement optimization for multi-user communications,'' {\em IEEE Trans.
		Commun.}, vol.~67, no.~8, pp.~5645--5658, 2019.
	
	\bibitem{sutton2018reinforcement}
	R.~S. Sutton and A.~G. Barto, {\em {Reinforcement Learning: An Introduction}}.
	\newblock MIT press, 2018.
	
	\bibitem{bjornson2010cooperative}
	E.~Bj{\"o}rnson, R.~Zakhour, D.~Gesbert, and B.~Ottersten, ``{Cooperative
		multicell precoding: Rate region characterization and distributed strategies
		with instantaneous and statistical CSI},'' {\em IEEE Trans. Signal Process.},
	vol.~58, no.~8, pp.~4298--4310, 2010.
	
	\bibitem{dulac2015deep}
	{G. Dulac-Arnold \emph{et al.}}, ``Deep reinforcement learning in large
	discrete action spaces,'' {\em \url{arXiv:1512.07679}}, 2015.
	
	\bibitem{hur2013millimeter}
	{S. Hur \emph{et al.}}, ``Millimeter wave beamforming for wireless backhaul and
	access in small cell networks,'' {\em IEEE Trans. Commun.}, vol.~61, no.~10,
	pp.~4391--4403, 2013.
	
	\bibitem{lillicrap2015continuous}
	{T. P. Lillicrap \emph{et al.}}, ``Continuous control with deep reinforcement
	learning,'' {\em in Proc. ICLR 2016, arXiv:1509.02971}.
	
\end{thebibliography}
\end{document}